# Permeation of Water Nanodroplets on Carbon Nanotubes Forests

Ygor M. Jaques and Douglas S. Galvao
Applied Physics Department, University of Campinas, Campinas, SP 13081-970, Brazil

## ABSTRACT

Fully atomistic molecular dynamics simulations were carried out to investigate how a liquid-like water droplet behaves when into contact with a nanopore formed by carbon nanotube arrays. We have considered different tube arrays, varying the spacing between them, as well as, different chemical functionalizations on the uncapped nanotubes. Our results show that simple functionalizations (for instance, hydrogen ones) allow tuning up the wetting surface properties increasing the permeation of liquid inside the nanopore. For functionalizations that increase the surface hydrophilicity, even when the pore size is significantly increased the droplet remains at the surface without tube permeation.

## INTRODUCTION

The wetting dynamics of surfaces is an important area of fluid dynamics, the wetting behavior many times determines the potential use of a specific material in technological applications [1]. Because of that, the engineering of texturized structures at the micro/nanoscopic level has been used to obtain surfaces with extraordinary properties, like rapid detachment [2], superhydrophilicity and superhydrophobicity [3]. Some of these properties can be very useful to applications concerning anti-icing, antifogging and self-cleaning materials. As an example, engineering patterned surfaces created by placing uniformly spaced pillars, represent an effective and feasible strategy to design materials with such properties [4,5].

In order to understand, at atomic level, how a liquid like water behaves when into contact with such materials, we have investigated through fully atomistic classical molecular dynamics (MD) simulations the permeation behavior of water nanodroplets into contact with patterned carbon nanotubes (CNT) forests (our patterned surface – see Figure 1). More specifically, we have considered different tube arrays, where we varied the spacing between tubes. We have also considered tubes with different edge functional groups (hydrogen and hydroxyl ones) in order to create different degrees of array wettability and/or hydrophilic or hydrophobic behaviors.

## THEORY

The MD simulations were carried out using the LAMMPS code [6]. Our system consisted of 28 (8,0) carbon nanotubes forming a regular pattern to emulate a nanoforest. The length of these nanotubes is 10 nm. All tube-dangling bonds were functionalized (passivated) with hydrogen or hydroxyl groups. It is important to stress that as we are using nanotubes of very small diameters (water molecules cannot enter them), we can investigate the permeation that occur only at the nanopores.

To model the behavior of a water nanodroplet near the surface of a nanopore (created by manipulating the nanotube spatial arrays), three different separations were considered at the middle of the nanoforest: 5, 10 and 20 Å. A representation of the initial configuration system is depicted in Figure 1.

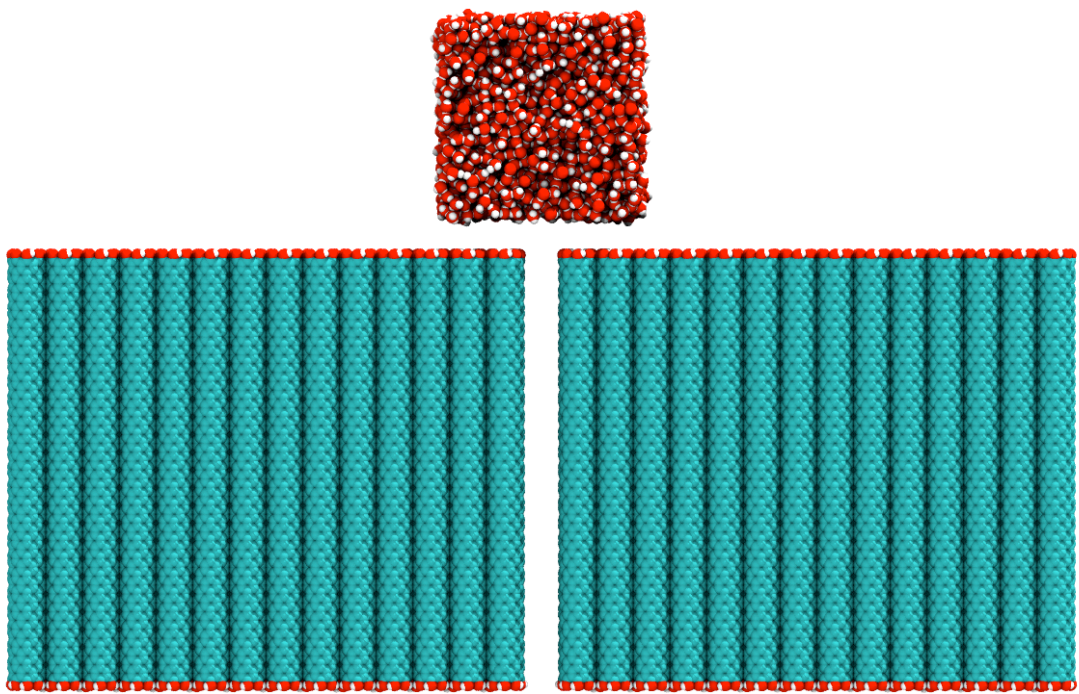

**Figure 1.** Initial configuration for the water nanodroplets dynamics placed above the nanopore surface.

The droplets were first equilibrated (at 300 K) with a Nosé-Hoover thermostat [7,8], keeping the center of mass (COM) fixed, assuming the expected spherical shapes. After that, we remove the COM restraint and the droplet start to move towards the nanoforest surface.

Two different functionalizations were considered at tube edges, hydrogenations and hydroxylations. The SPC/E model [9] was used for water and the C-O parameter interactions were obtained from [10]. Other nonbonded interactions were obtained from the standard CHARMM forcefield [11].

The spatial density distribution of the final droplet configuration is obtained by dividing the simulation cell into boxes of dimensions 0.5 x 0.5 x 50 Å$^3$. The density was calculated considering a MD time run of 200 ps with time steps of 2.0 fs and temperatures always at 300 K. This methodology approach has been proved to be effective in the study of nanodroplets [11].

**DISCUSSION**

We start analyzing how the different functionalizations affect the droplet dynamics behavior. Our simulations showed that hydrogen functionalizations tend to improve the hydrophobic character of the nanoforest, while the hydroxyl functionalizations have the opposite effect, i. e., they increase the hydrophilic character. This can be seen in Figure 2, where we present representative MD snapshots of the equilibrium configurations for the two-functionalization cases. As we can see from this Figure, the droplet is not widely spread on the hydrogenated surface and a significant droplet part permeates through the nanopore. However, for the hydroxylated surface the droplet is spread over almost all-accessible area and the permeated part is much smaller in comparison to the hydrogenated cases. However, as mentioned above, for both cases, there is some degree of capillarity (tube permeation), which stabilizes after 2 ns.

For the H-functionalization case, it is very clear that there is a tendency of the liquid to go inside the pore, even with the nanotube surface being also hydrophobic. It is interesting how very simple functionalizations can tune the behavior of the liquid inside a nanopore, and can assume that for more complex functionalizations (with more hydrocarbon chains) this behavior can be increased.

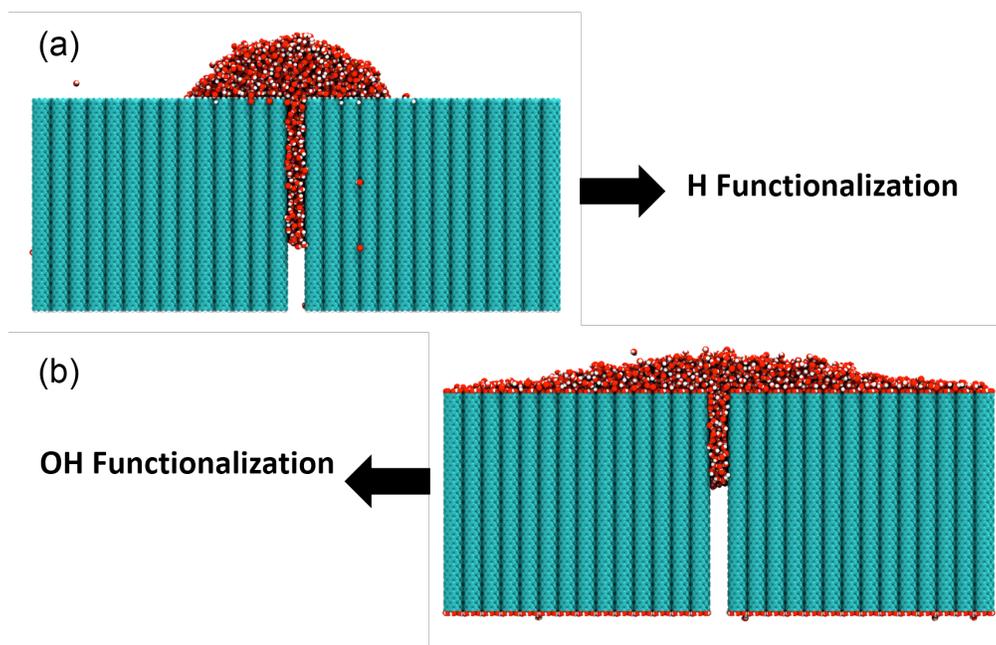

**Figure 2.** MD snapshots of the equilibrium configurations for Hydrogenation (a) and OH (b) functionalizations for a separation nanopore of 5 Å created in the middle of the nanoforest.

One important magnitude to character the wetting behavior is contact angle values. From the MD results we calculated these values, as well as the density profile of the equilibrated droplets. An angle of 74° was obtained for H-functionalization (Figure 3) and for OH-functionalization this value decreased to as low as 15°. Due to surface pore homogeneity, we can expect the equilibrium pore contact angle values to be close to zero degrees. However, the exact contact angle value is difficult to estimate because the pores are too small to provide a clear liquid-vacuum profile to curve fitting.

We also investigated how the pore size affects the droplet dynamics (spreading and permeation). We considered 3 sizes: 5, 10 and 20 Å. For the H-functionalization cases, increasing the pore size increases the droplet permeation, as the liquid easier flows through the pore (Figure 4). For the largest spacing (20 Å), we can see some stabilization of the water molecules near the top surface, even if the liquid has more space to flow inside. This could be understood as a consequence of the tube internal surface hydrophobicity (no good affinity with the liquid).

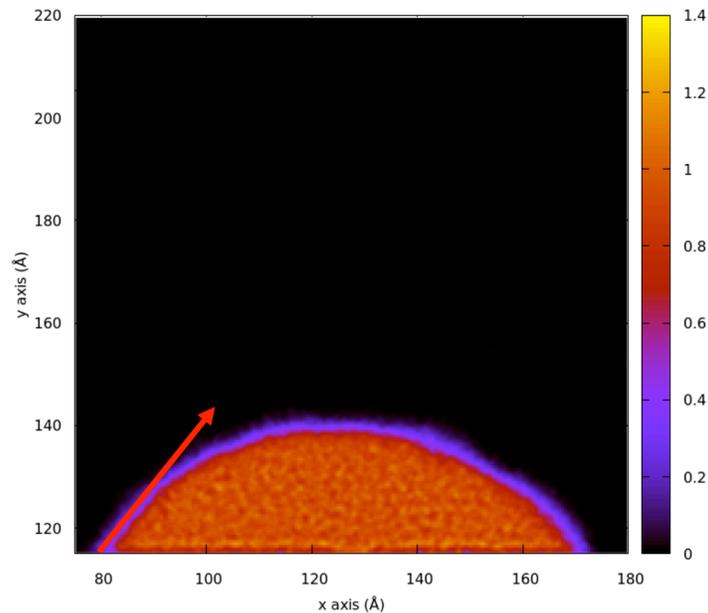

**Figure 3.** Superficial density (arbitrary units) of the water nanodroplet formed on the H-functionalized. Results for a 5 Å nanopore. The arrow indicates the calculated contact angle value (measure on clockwise direction).

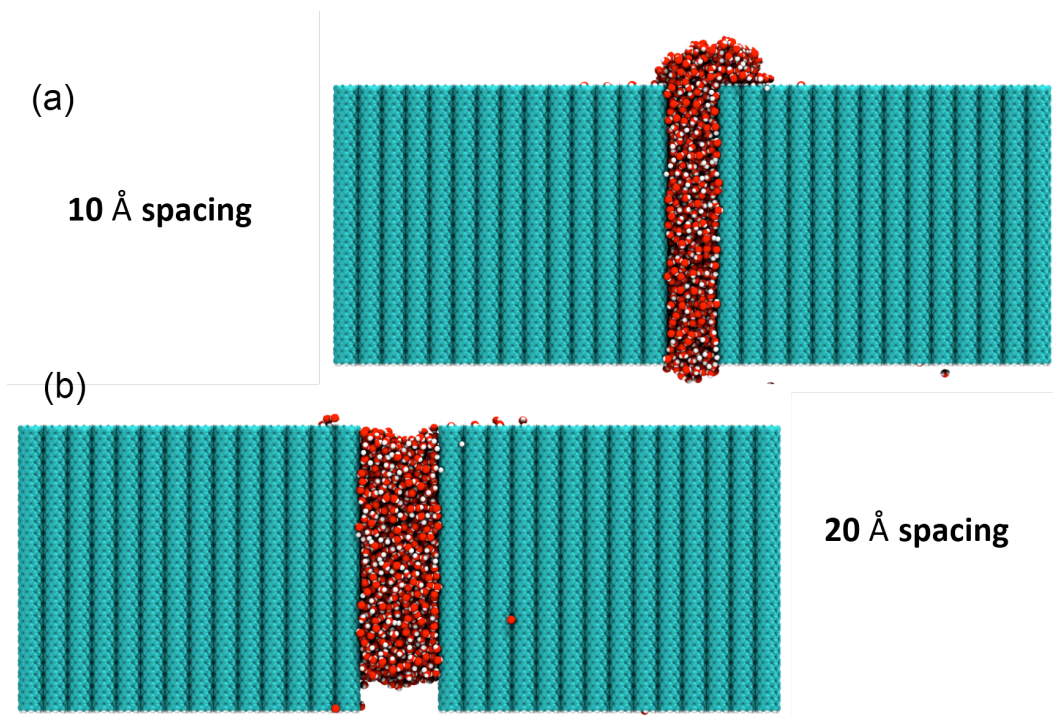

**Figure 4.** Representative MD snapshots for stable droplet configurations. H-functionalization results for pores of 10 (a) and 20 Å (b), respectively.

For the OH-functionalization cases, a quite different behavior occurs as the hydrophilic coating on the extremities of the tubes causes a very strong interaction between liquid and the top surface (Figure 5). Even with a spacing of 20 Å the liquid remains widely spread over the top of the nanoforest with smaller permeation in relation to the hydrogenated cases.

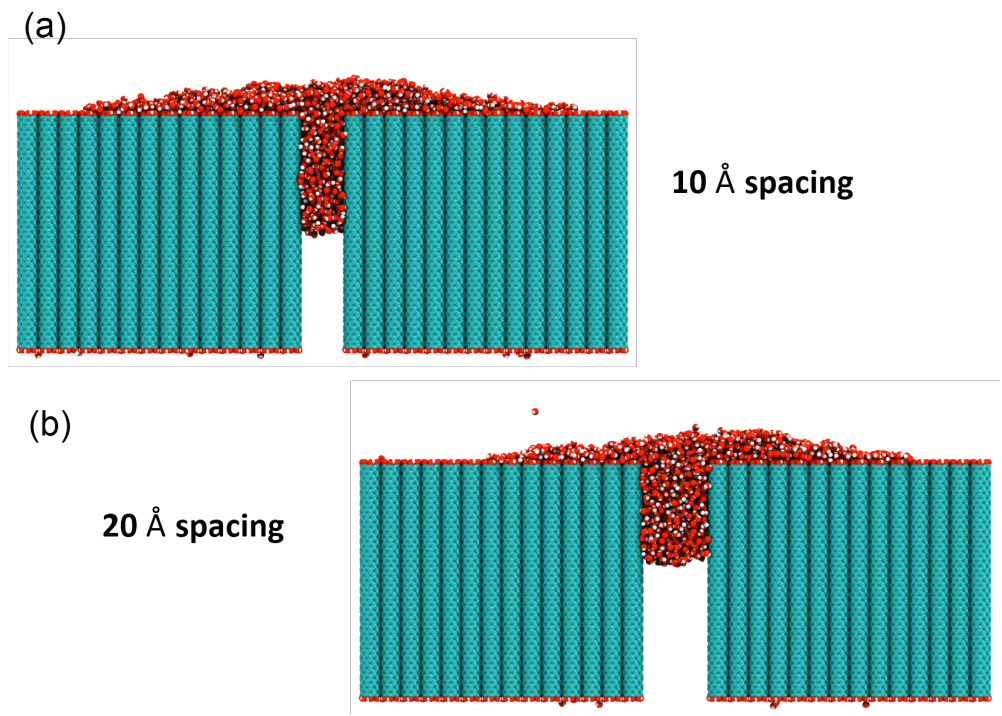

**Figure 5.** Representative MD snapshots for stable droplet configurations. OH-functionalization results for pores of 10 (a) and 20 Å (b), respectively.

**CONCLUSIONS**

We have investigated, through fully atomistic molecular dynamics (MD) simulations the wetting behavior of water-like droplets deposited on nanopores formed by functionalized (H and OH groups) carbon nanotube arrays (nanoforests).

Our results show that using simple chemical functionalizations it is possible to tune the wetting response (in terms of droplet spreading and permeation) of the nanotube forests, in spite of the tube internal surface hydrophobicity. For the H-functionalization cases, the droplet does not exhibit significant spreading and increasing the pore size increases the droplet permeation. For the OH-functionalization cases, the droplets are widely spread on the forest surface and even for the largest spacing investigated here (20 Å) the droplet permeation is not very extensive. The reasons for this differentiated behavior were discussed in details in the above section.

As our structures are very small (evaporation effects are still important and droplet fragmentation involves sometimes only a few water molecules) a direct comparison with classical wetting theories is not possible. These results are suggestive that they are direct consequences of spatial variations at atomic size scale, thus it could not be captured by classical theories. We need further tests to confirm this hypothesis. These results can be exploited in the design of nanotexturized surfaces by top-down or bottom-up techniques [12].

ACKNOWLEDGMENTS

This work was supported in part by the Brazilian Agencies CAPES, CNPq and FAPESP. The authors also thank the Center for Computational Engineering and Sciences at Unicamp for financial support through the FAPESP/CEPID Grant # 2013/08293-7.

**REFERENCES**

1. L. Duta, A.C. Popescu, I. Zgura, N. Preda and I.N. Mihailescu, Wettability of Nanostructured Surfaces, Wetting and Wettability, edited by Dr. Mahmood Aliofkhazraei (Intech, 2015) p. 207.
2. Y. Liu, L. Moevius, X. Xu, T. Qian, J. M. Yeomans, and Z. Wang, Nat. Phys. **10**, 515 (2014).
3. Z. Wang, M. Elimelech, and S. Lin, Environ. Sci. Technol. (2016).
4. Y. Nonomura, T. Tanaka, and H. Mayama, Langmuir **0**, null (n.d.).
5. L. Zhang and D. Resasco, Langmuir **25(8)**, 4792 (2009).
6. S. Plimpton, J. Comput. Phys. **117**, 1 (1995).
7. S. Nosé, J. Chem. Phys. **81**, 511 (1984).
8. W. G. Hoover, Phys. Rev. A **31**, 1695 (1985).
9. H. J. C. Berendsen, J. R. Grigera, and T. P. Straatsma, J. Phys. Chem. **91**, 6269 (1987).
10. R. L. Jaffe, P. Gonnet, T. Werder, J. H. Walther, and P. Koumoutsakos, Mol. Simul. **30**, 205 (2004).
11. A. D. MacKerell, D. Bashford, M. Bellott, R. L. Dunbrack, J. D. Evanseck, M. J. Field, S. Fischer, J. Gao, H. Guo, S. Ha, D. Joseph-McCarthy, L. Kuchnir, K. Kuczera, F. T. K. Lau, C. Mattos, S. Michnick, T. Ngo, D. T. Nguyen, B. Prodhom, W. E. Reiher, B. Roux, M. Schlenkrich, J. C. Smith, R. Stote, J. Straub, M. Watanabe, J. Wiórkiewicz-Kuczera, D. Yin, and M. Karplus, J. Phys. Chem. B **102**, 3586 (1998).
11. Y. M. Jaques, G. Brunetto and D. S. Galvao, *MRS Adv.* **1**, 675 (2016).
12. Y. M. Jaques and D. S. Galvao, *to be published*.